\documentclass{PoS}
\usepackage{url}
\def\spose#1{\hbox to 0pt{#1\hss}}
\def\ltapprox{\mathrel{\spose{\lower 3pt\hbox{$\mathchar"218$}}
 \raise 2.0pt\hbox{$\mathchar"13C$}}}
\def\gtapprox{\mathrel{\spose{\lower 3pt\hbox{$\mathchar"218$}}
 \raise 2.0pt\hbox{$\mathchar"13E$}}}
\def\inapprox{\mathrel{\spose{\lower 3pt\hbox{$\mathchar"218$}}
 \raise 2.0pt\hbox{$\mathchar"232$}}}
 
\def\CO{{\cal O}}
\def\bar{\overline}

\def\beq{\begin{equation}}
\def\eeq{\end{equation}}

\def\bea{\begin{eqnarray}}
\def\eea{\end{eqnarray}}

\title{Pseudoscalar decay constants,
light-quark masses, and $B_K$ from
mixed-action lattice QCD}

\ShortTitle{Decay constants, quark masses, and $B_K$ from
mixed-action lattice QCD}

\author{Jack Laiho\thanks{Funded by STFC and the Scottish Universities Physics Alliance.}\\
        Department of Physics and Astronomy, University of Glasgow, Glasgow, Scotland, UK\\
        E-mail: \email{jlaiho@fnal.gov}}

\author{\speaker{Ruth S. Van de Water}\thanks{This manuscript has been authored by employees of Brookhaven Science Associates, LLC under Contract No. DE-AC02-98CH10886 with the U.S. Department of Energy.  Computations for this work were carried out with resources provided by the USQCD Collaboration, the Argonne Leadership Computing Facility, and the New York Center for Computational Sciences, which are funded by the Office of Science of the U.S. Department of Energy and by New York State.}\\
        Physics Department, Brookhaven National Laboratory, Upton, New York, USA\\
        E-mail: \email{ruthv@bnl.gov}}

\abstract{We present updated results for the leptonic decay constants $f_\pi$ and $f_K$, the light $u$, $d$, and $s$-quark masses, and the neutral kaon mixing parameter $B_K$ from mixed-action lattice simulations with staggered sea quarks and domain-wall valence quarks. We use the publicly-available 2+1 flavor MILC asqtad-improved staggered gauge configurations with multiple light sea-quark masses and three lattice spacings, and compute the kaon mixing matrix element with several partially-quenched valence-quark masses. We then extrapolate to the physical light-quark masses and the continuum using partially-quenched chiral perturbation theory formulated for mixed-action lattice simulations. For $B_K$ we match the lattice four-fermion operator to the continuum using the nonperturbative method of Rome-Southampton. Our new results benefit from two significant improvements over our published work:  (1) we have added a third lattice spacing of a$\approx$0.06 fm to better control the continuum extrapolation, and (2) we have implemented a new lattice renormalization scheme (the RI/SMOM$_{\gamma_\mu}$ scheme developed by Sturm {\it et al.}) that suppresses chiral-symmetry breaking and other infrared effects and, in practice, also shrinks the size of the 1-loop perturbative coefficient needed to match to the continuum $\bar{\textrm{MS}}$ scheme. When combined with the use of volume-averaged momentum sources and twisted-boundary conditions, this significantly reduces the systematic uncertainty in the renormalization factor $Z_{B_K}$.}  

\FullConference{ The XXIX International Symposium on Lattice Field Theory - Lattice 2011\\
July 10-16, 2011\\
Squaw Valley, Lake Tahoe, California}

\begin{document}

\vspace{-10mm}
%%%%%%%%%%%%%%%
\section{Motivation}
%%%%%%%%%%%%%%%

Lattice-QCD calculations of pseudoscalar decay constants, light-quark masses, and other kaon weak matrix elements are important ingredients for understanding the Standard Model and for constraining physics beyond the Standard Model.  For example, the $u$, $d$, and $s$-quark masses are parametric inputs to calculations of Standard Model and new physics processes.  The ratio of pseudoscalar decay constants $f_K/f_\pi$, when combined with experimental measurements of the leptonic decay rates, allows a precise determination of the ratio of CKM matrix elements $|V_{ud}|/|V_{us}|$~\cite{Marciano:2004uf}.  The neutral kaon mixing parameter $B_K$, when combined with experimental measurements of indirect $CP$-violation in the kaon sector ($\varepsilon_K$) constrains the apex of the CKM unitarity triangle.  Finally, penguin-dominated $K\to\pi\pi$ and $K\to\pi\nu\bar{\nu}$ decays may be particularly sensitive probes of new physics once lattice weak matrix element calculations are sufficiently precise.

\begin{figure}
\begin{center}
\includegraphics[width=0.65\linewidth]{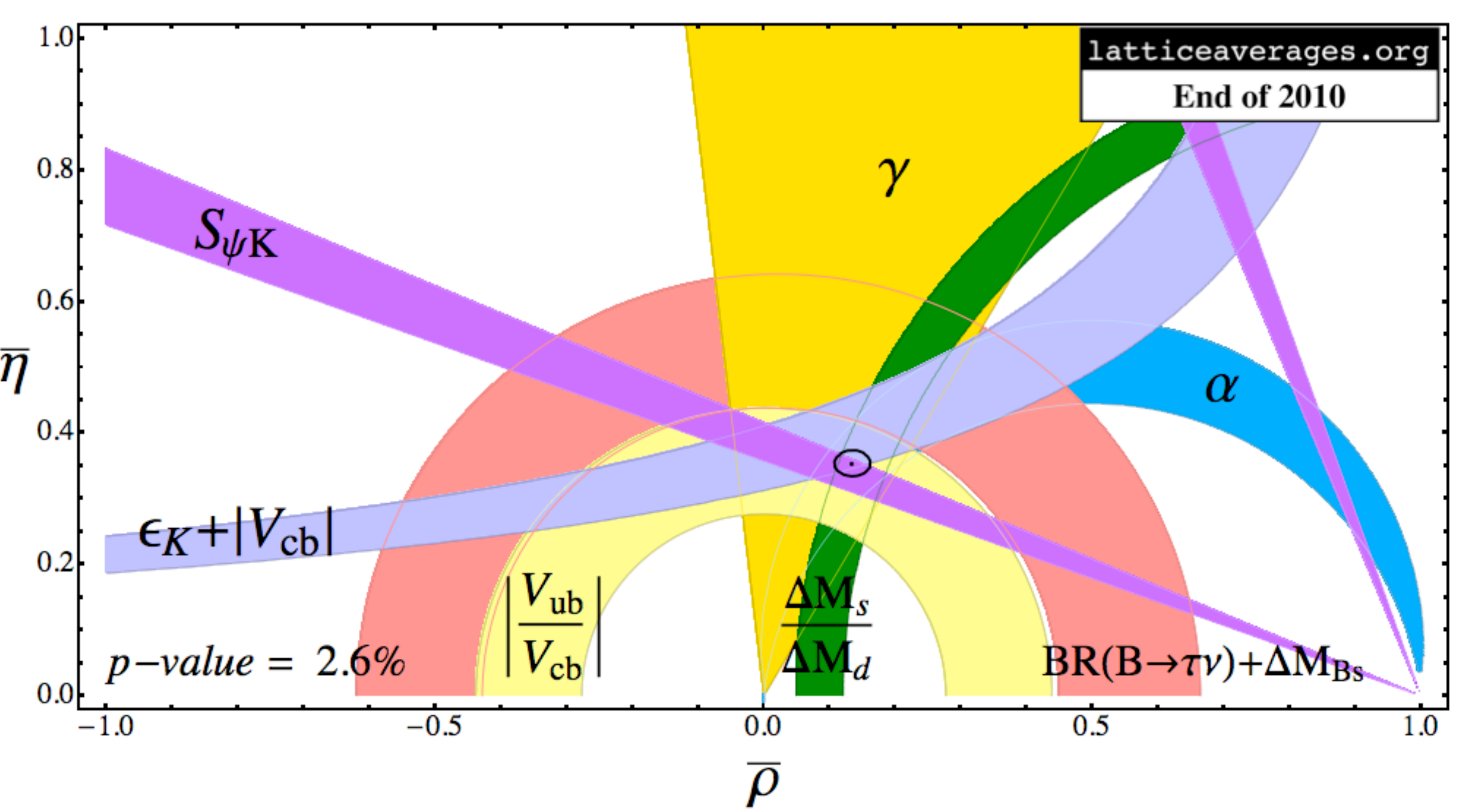}
\vspace{-2mm}
\caption{Global fit of the CKM unitarity triangle~\cite{Laiho:2009eu}.}\label{utfit-full.pdf}
\end{center}
\end{figure}

%Over the past decade, 
Lattice QCD has played a key role in establishing that the CKM paradigm of $CP$-violation describes experimental observations at the $\sim\!\!10\%$ level~\cite{Antonelli:2009ws}.  Many new-physics scenarios, however, predict new interactions between quarks and non-standard $CP$-violating phases.  Given sufficient theoretical and experimental precision, these would lead to inconsistencies between measurements that are expected to be the same within the Standard Model CKM framework.  Recent improvements in lattice weak matrix element calculations, especially of $B_K$ and the $B$-mixing $SU(3)$-breaking ratio $\xi$, have revealed a $\sim\!\!3\sigma$ tension that may indicate the presence of a non-Standard Model source of $CP$-violation (see Fig.~\ref{utfit-full.pdf})~\cite{NPCKM}.  Given this tension, it is crucial to continue precision studies of kaon physics using multiple methods including our mixed-action approach.

%%%%%%%%%%%%%%%
\section{Overview of mixed-action calculation}
%%%%%%%%%%%%%%%

Table~\ref{tab:MILC} shows the parameters of our numerical mixed-action lattice simulations.  We analyze the 2+1 flavor asqtad-improved staggered gauge configurations generated by the MILC Collaboration~\cite{Aubin:2004fs}.  Use of this large suite of ensembles with multiple spatial volumes $\sim$ (2.5 -- 4 fm)$^3$, multiple light sea-quark masses $\sim m_s/10$ -- $m_s$, and three lattice spacings $\sim$ 0.06 -- 0.12~fm gives us good control over the systematic uncertainties associated with the chiral-continuum extrapolation.   %In particular, our lightest (taste-pseudoscalar) sea-sea pion has $m_{\pi,5} = 240$~MeV, and on the ``superfine" $a \sim 0.06$ fm ensembles, the heaviest (taste-singlet) sea-sea pion is also quite light with $m_{\pi,I} = 270$~MeV.  
We generate domain-wall valence quarks at several partially-quenched masses $\sim m_s/10$ -- $m_s$.  We apply HYP-smearing~\cite{Hasenfratz:2001hp} to the valence domain-wall action in order to reduce the size of explicit chiral symmetry breaking~\cite{Renner:2004ck} ($m_{\rm res}~\approx~3$~MeV at our coarsest lattice spacing and is $\sim 30$ times smaller at our finest lattice spacing).  The use of domain-wall valence quarks makes the chiral-continuum extrapolation more continuum-like~\cite{MAChPT,BK_MAChPT} and simplifies the nonperturbative operator matching via the method of Rome-Southampton~\cite{Martinelli:1994ty}.  Hence our mixed-action approach is well-suited for weak matrix element calculations, as we show empirically in the next section.

\begin{table}\begin{center}
\begin{tabular}{ccccccccc} \\ \hline\hline
$a$(fm) & $L^3 \times T$  & $m_l$ & $m_h$ &  $m_{\rm val.}^{\rm dwf}$ & \# configs.   \\[0.5mm] \hline
\textbf{$\approx$ 0.06} & \textbf{64$^3 \times$ 144} & \textbf{0.0018} & \textbf{0.018} & \textbf{0.0026, 0.0469, 0.0108, 0.033} & \textbf{96} \\
\textbf{$\approx$ 0.06} & \textbf{48$^3 \times$ 144} & \textbf{0.0036} & \textbf{0.018} & \textbf{0.0036, 0.0072, 0.0108, 0.033} & \textbf{128} \\
\hline
\textbf{$\approx$ 0.09} & \textbf{40$^3 \times$ 96} & \textbf{0.0031} & \textbf{0.0031} & \textbf{0.004, 0.0124, 0.0186, 0.046} & \textbf{102} \\
{$\approx$ 0.09} & {$40^3 \times 96$} & {0.0031} & {0.031} & {0.004, 0.0124, 0.0186, 0.046} & {150} \\
{$\approx$ 0.09} & {$28^3 \times 96$} & {0.0062} & {0.031} & 0.0062, 0.0124, 0.0186, 0.046 & 374 \\
\textbf{$\approx$ 0.09} & \textbf{28$^3 \times$ 96}\ & \textbf{0.0093} & \textbf{0.031} & \textbf{0.0062, 0.0124, 0.0186, 0.046} & \textbf{198} \\
{$\approx$ 0.09} & {$28^3 \times 96$} & {0.0124} & {0.031} & 0.0062, 0.0124, 0.0186, 0.046 & {198} \\
{$\approx$ 0.09} & {$28^3 \times 96$} & {0.0062} & {0.0186} & 0.0062, 0.0124, 0.0186, 0.046 & 160 \\
\hline
\textbf{$\approx$ 0.125} & \textbf{32$^3 \times$ 64} & \textbf{0.005} & \textbf{0.005} & \textbf{0.007, 0.02, 0.03, 0.05} & \textbf{175} \\
{$\approx$ 0.125} & {$24^3 \times 64$} & {0.005} & {0.05} & 0.007, 0.02, 0.03, 0.05, 0.065 & 216 \\
{$\approx$ 0.125} & {$20^3 \times 64$} & {0.007} & {0.05} & 0.01, 0.02, 0.03, 0.04, 0.05, 0.065 & 268 \\
{$\approx$ 0.125} & {$20^3 \times 64$} & {0.01} & {0.05} & 0.01, 0.02, 0.03, 0.05, 0.065 & 220 \\
{$\approx$ 0.125} & {$20^3 \times 64$} & {0.02} & {0.05} & 0.01, 0.03, 0.05, 0.065 & 117 \\
{$\approx$ 0.125} & {$20^3 \times 64$} & {0.01} & {0.03} & 0.01, 0.02, 0.03, 0.05, 0.065 & 160\\
\hline\hline
\end{tabular}\caption{Sea-quark ensembles and valence-quark masses used to obtain the preliminary results presented in this work.  Ensembles shown in \textbf{bold} are new since our 2009 $B_K$ publication~\cite{Aubin:2009jh}.}\label{tab:MILC}\end{center}\end{table}

%%%%%%%%%%%%%%%
\section{Testing the mixed-action method: decay constants and quark masses}
%%%%%%%%%%%%%%%

\begin{figure}
\begin{center}
\includegraphics[height=0.52\linewidth]{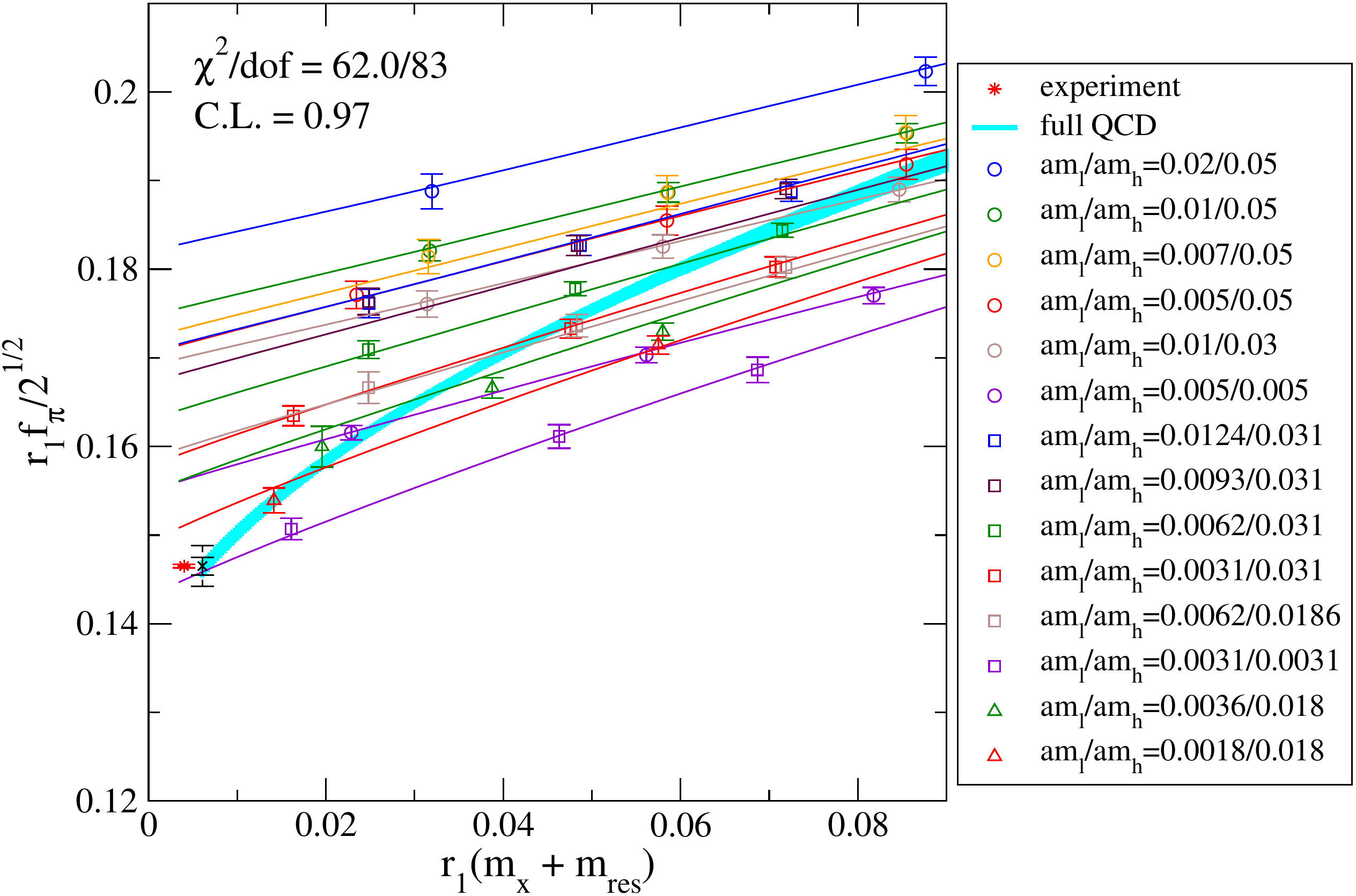} 
\caption{Chiral-continuum extrapolation of the pseudoscalar decay constant.  Only data points with degenerate valence-quark masses are shown, but the fit includes all nondegenerate points.  The circles denote ``coarse" ($a \approx 0.12$~fm) data, the squares denote ``fine" ($a \approx 0.09$~fm) data, and the triangles denote ``superfine" data.  The cyan band shows the continuum full QCD curve.  The black ``$\times$" shows $f_\pi$ at the physical $ud$ quark mass, where the inner solid error bar is statistical and the outer dashed error bar is the total systematic.  For comparison, the red star shows the experimental result for $f_\pi$, with which we are in good agreement.}\label{fig:fpi}
\end{center}
\end{figure}

%%
%\beq
%	f_P = \frac{A_{WP}}{\sqrt{A_{WW}}} \frac{\sqrt{2}(m_x+m_y+2m_{res})}{m_\pi^{3/2}} \,,
%\eeq
%%
%where the subscripts $WW$ and $WP$ denote wall-source, wall-sink and wall-source, point-sink two point correlation functions, respectively.  
We compute the decay constant using a Ward identity to relate the axial current to the pseudoscalar density;  this method has the advantage that the renormalization factor is unity up to corrections of $\CO(am_{\rm res})\sim10^{-3}$--$10^{-5}$.  Once we have the decay constant for all combinations of valence- and sea-quark masses, we perform a combined chiral-continuum extrapolation via a simultaneous correlated fit of the full data set using NLO $SU(3)$ mixed-action $\chi$PT~\cite{MAChPT} supplemented with higher-order analytic terms.  Because the kaon mass is so heavy and our data is so precise, these terms are needed to interpolate about the strange-quark mass and obtain good confidence levels.  We correct the numerical data for the known one-loop finite volume effects, and estimate the systematic uncertainty due to the chiral-continuum extrapolation by varying the fit function.   As shown in Fig.~\ref{fig:fpi}, the value of $f_\pi$ in the continuum limit at the physical quark masses agrees with experiment.  Our preliminary results for the pseudoscalar decay constants and their ratio are
\begin{eqnarray}
f_\pi &=& 130.53(0.87)_{\rm stat}(1.68)_{\chi {\rm PT}}(0.80)_{\rm FV}(0.93)_{r_1}(0.25)_{m_q} ~\textrm{MeV} \label{eq:fpi} \\
f_K &=& 156.8(1.0)_{\rm stat}(1.1)_{\chi {\rm PT}}(0.6)_{\rm FV}(0.8)_{r_1}(0.8)_{m_q}  ~\textrm{MeV} \\
f_K/f_\pi &=& 1.202(0.011)_{\rm stat}(0.009)_{\chi {\rm PT}}(0.008)_{\rm FV}(0.002)_{r_1}(0.005)_{m_q} \label{eq:fK_fpi}  \,,
\end{eqnarray}
where the errors are from statistics, the chiral-continuum extrapolation, finite volume effects, the scale uncertainty, and the light-quark mass uncertainties, respectively.  We convert lattice quantities into physical units with the value of the scale $r_1$ obtained by the HPQCD collaboration from several quantities including pseudoscalar decay constants and $\Upsilon$ splittings~\cite{Davies:2009tsa}.  

We determine the bare average $ud$ and $s$ domain-wall valence-quark masses by requiring that the pion and kaon masses in the continuum at these quark masses take their experimental values~\cite{Nakamura:2010zzi}.  We compute the quark-mass renormalization factor $Z_m = 1/Z_S$ using a partly-nonperturbative approach in which we combine a nonperturbative determination of $Z_A$ with a one-loop tadpole-improved lattice perturbation theory~\cite{Lepage:1992xa} calculation of $(1-Z_A/Z_S)$.  Because the ratio $Z_A/Z_S$ is close to unity, the size of the one-loop correction is small; nevertheless, the truncation error is still the largest source of uncertainty in $Z_m$.  We multiply the bare-quark masses by $Z_m$ to obtain the continuum $\bar{\textrm{MS}}$ masses;  our preliminary results are
\begin{eqnarray}
m_s^{\overline{\textrm{MS}}(\textrm{2 GeV})} &=& 94.2(1.4)_{\rm stat}(3.2)_{\rm sys}(4.7)_{\rm match} ~\textrm{MeV} \label{eq:ms}\\
m_{ud}^{\overline{\textrm{MS}}(\textrm{2 GeV})} &=& 3.31(0.07)_{\rm stat}(0.20)_{\rm sys}(0.17)_{\rm match} ~\textrm{MeV}\\
m_d^{\overline{\textrm{MS}}(\textrm{2 GeV})} &=& 4.73(0.09)_{\rm stat}(0.27)_{\rm sys}(0.24)_{\rm match} ~\textrm{MeV} \\
 m_u^{\overline{\textrm{MS}}(\textrm{2 GeV})}&  = &1.90(0.08)_{\rm stat}(0.21)_{\rm sys}(0.10)_{\rm match} ~\textrm{MeV}\\
\left( m_s/m_{ud} \right) &=&  28.4(0.5)_{\rm stat}(1.3)_{\rm sys} \qquad \left( m_u/m_d \right) = 0.401(0.013)_{\rm stat}(0.045)_{\rm sys} \label{eq:mu_md} \,,
\end{eqnarray}
where the errors are from statistics, all systematics except for renormalization, and renormalization, respectively.  Although we do not include isospin-breaking in our lattice simulations, we obtain $m_u$ and $m_d$ separately following the method adopted by MILC~\cite{Aubin:2004fs}, which uses the difference between the $K^+$ and $K^0$ meson masses and continuum estimates for the violation of Dashen's theorem.

Our preliminary determinations of the decay constants and light-quark masses
%in Eqs.~(\ref{eq:fpi})--(\ref{eq:fK_fpi}) and the quark masses in Eqs.~(\ref{eq:ms})--(\ref{eq:mu_md})
agree well with $N_f=2+1$ lattice QCD calculations by other collaborations~\cite{ALL}  (see also the recent review by Wittig~\cite{HWittig}).  This gives us confidence that the mixed-action method is reliable and can be used to obtain $B_K$ and $K\to\pi\pi$ matrix elements precisely with controlled systematic uncertainties.

%%%%%%%%%%%%%%%
\section{The neutral kaon mixing parameter: $B_K$}
%%%%%%%%%%%%%%%

In these proceedings we focus on improvements to our published calculation of $B_K$ in Ref.~\cite{Aubin:2009jh}.
Since 2009 we have addressed the two largest sources of uncertainty in our earlier work:
\begin{enumerate}
\item \textit{The chiral-continuum extrapolation:}
We have added ensembles with larger volumes and lighter pions to shorten the chiral extrapolation.  We have also added
data at a third finer lattice spacing to reduce taste-breaking and shorten the continuum extrapolation.
\item \textit{The $\Delta S=2$ operator renormalization:}
We now compute the renormalization factor $Z_{B_K}$ using volume-averaged momentum sources to reduce statistical errors and twisted boundary conditions to eliminate scatter from $O(4)$ symmetry breaking~\cite{Arthur:2010ht}.  We also now compute $Z_{B_K}$ with non-exceptional kinematics to reduce chiral symmetry breaking~\cite{Sturm:2009kb}.
\end{enumerate}

We compute $Z_{B_K}$ via the nonperturbative method of Rome-Southampton~\cite{Martinelli:1994ty}.  We use an intermediate lattice scheme with
symmetric momentum $p_1^2 = p_2^2 = (p_1 - p_2)^2$ developed by Sturm {\it et al.}~\cite{Sturm:2009kb}; this suppresses chiral symmetry breaking and
leads to a more convergent perturbative series for the conversion factor between the intermediate lattice
RI/(S)MOM scheme and the continuum $\bar{\textrm{MS}}$ scheme.
Specifically, we use the RI/SMOM$_{\gamma_\mu}$ scheme (which has the same projectors as the RI/MOM scheme), for which the 1-loop correction to the continuum $\bar{\textrm{MS}}$ scheme at 2~GeV is $\sim$ 4 times smaller than for the standard RI/MOM scheme~\cite{Aoki:2010pe}.
Because the RI/SMOM$_{\gamma_\mu}$ scheme is defined for massless quarks,  we compute the bilinear and four-fermion vertex functions for several values of the valence, light, and strange sea-quark masses and extrapolate to the chiral limit ${m_{\rm val.}^{\rm dwf}, m_{l}, m_h}\to0$.  The largest source of uncertainty in our calculation of $Z_{B_K}$ is  the residual perturbative truncation error from conversion to the continuum $\bar{\textrm{MS}}$ scheme.  We estimate this in several ways including the size of 1-loop term (0.5\%), comparison with lattice perturbation theory (2.0\%), and the residual slope in $(r_1p)^2$ (2.2\%), and choose the largest error estimate to be conservative.

%Reduction in chiral symmetry breaking
%Compare vector and axial-vector bilinear vertex functions
%With non-exceptional momenta $\Lambda_V \approx \Lambda_A$ in the chiral limit at the sub-percent level

%\begin{figure}
%\begin{center}
%\includegraphics[width=0.45\linewidth]{grace/fpi_2pi_ratio.pdf} $\qquad$
%\includegraphics[width=0.45\linewidth]{grace/fK_2pi_ratio.pdf}
%%\vspace{-6mm}
%\caption{$\Lambda_V$ versus $\Lambda_A$ plot here.}\label{fig:LmdV_m_LmdA}
%\end{center}
%\end{figure}

\begin{figure}
\begin{center}
\includegraphics[height=0.52\linewidth]{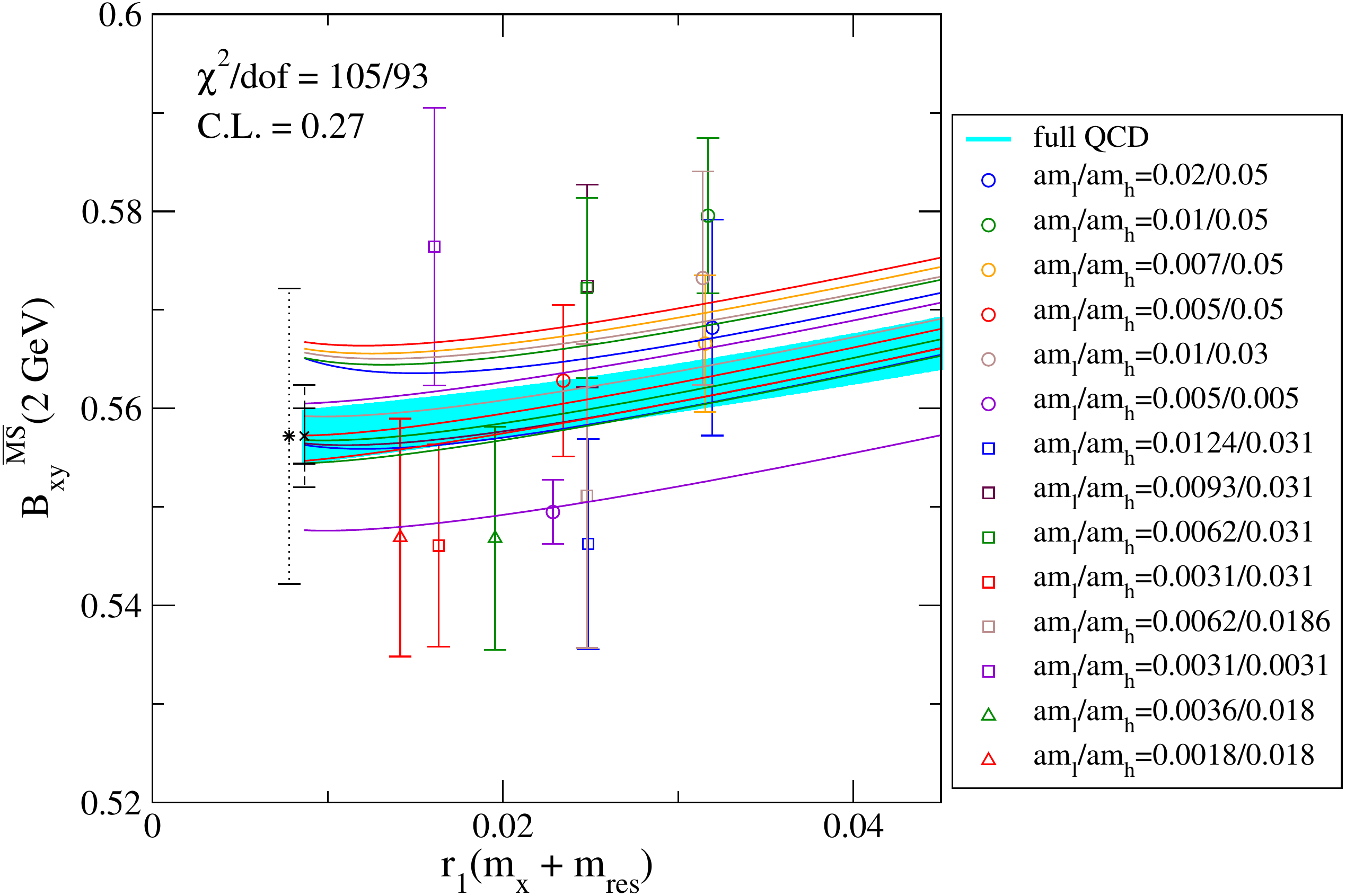} 
\caption{Chiral-continuum extrapolation of $B_K$.  The fit includes all available data, but only the nondegenerate data points in which one valence-quark mass is close to the strange quark and the other valence-quark mass is the lightest on that ensemble are shown.  The black ``$\times$" shows $B_K$ in the continuum at the physical $d$ and $s$ quark masses, where the inner solid error bar is statistical and the outer dashed error bar includes the chiral extrapolation systematic.  The star with dotted error bar, slightly offset, shows the total error in $B_K$.}\label{fig:BK}
\end{center}
\end{figure}

We compute the $B_K$ matrix element using symmetric and antisymmetric linear combinations of periodic and antiperiodic B.C.
quark propagators to minimize finite temporal size effects.  We obtain $\sim$0.5-1.5\% statistical errors in the lattice matrix element.
We use the same approach for the chiral-continuum extrapolation as for the pseudoscalar masses and decay constants.  We fit the renormalized $B_K$ data to the NLO $SU(3)$ mixed-action
$\chi$PT expression~\cite{BK_MAChPT}  supplemented by higher-order analytic terms (see Fig.~\ref{fig:BK}).
%Figure~\ref{fig:BK} shows the chiral-continuum extrapolation of $B_K$;  our data displays only a mild lattice-spacing dependence.
We estimate the systematic uncertainty due to the chiral extrapolation by varying the fit
function in several ways including using an analytic function without chiral logarithms, adding higher-order analytic and logarithm terms, and changing the value of $f_\pi$ in the coefficient of the 1-loop chiral logarithms.  We conservatively take the largest difference from our preferred fit as the error.  

We obtain the following preliminary result for $B_K$ in the $\bar\textrm{MS}$ scheme at 2~GeV:
%Table~\ref{tab:BK_Err} presents our preliminary error budget for $B_K$;  we obtain a 2.8\% total uncertainty:
%
\beq
	%B^{\overline{\textrm{MS}}\textrm{(2 GeV)}}_K = 0.5572(28)_\textrm{stat}(151)_\textrm{sys}\,,
	B^{\overline{\textrm{MS}}\textrm{(2 GeV)}}_K = 0.5572(28)_{\rm stat} (45)_{\chi {\rm PT}} (33)_{\rm FV} (39)_{r_1,m_q} (6)_{\rm EM} (134)_{\rm match}\,,
\eeq
where the error labels are the same as in previous equations; the total uncertainty is 2.8\%.    The largest contribution to the error is from the renormalization factor;  we aim to improve the uncertainty in $Z_{B_K}$ due to the chiral extrapolation with additional data before publication.
Ultimately, the dominant truncation error  in $Z_{B_K}$ may be reduced with a 2-loop continuum perturbative QCD calculation.  Our new result is $\sim1.1\sigma$ higher than our published value, and agrees well with calculations from several other lattice methods~\cite{BK} (see also the recent review by Mawhinney~\cite{BMawhinney}).

%\begin{table}
%\begin{center}
%\caption{Preliminary error budget for $B_K$.  Errors are shown as percentages.}\label{tab:BK_Err}
%\begin{tabular}{lr} \\ \hline\hline
%uncertainty & $B_K$  \\[0.5mm] \hline
%statistics & 0.5\% \\
%chiral and continuum extrapolation $\qquad$ &  0.8\%  \\
%renormalization factor & 2.4\%  \\
%scale and quark-mass uncertainties & 0.7\%  \\
%finite volume errors & 0.6\%   \\
%electromagnetic effects & 0.1\%  \\
%\hline
%total & 2.8\%  \\
%\hline\hline
%\end{tabular}
%\end{center}
%\end{table}

%%%%%%%%%%%%%%%
\section{Summary and outlook}
%%%%%%%%%%%%%%%

Mixed-action lattice QCD simulations are well-suited to the calculation of weak matrix elements, as shown by agreement with experiment and independent lattice results for pseudoscalar decay constants and light-quark masses, and also by our precise determination of $B_K$.  Given this and other recent improvements in lattice QCD calculations of $B_K$, it is no longer the dominant source of uncertainty in the $\varepsilon_K$ band.  Hence, although recent $B_K$ results slightly reduce the tension in the global CKM unitarity triangle fit, the tension remains at the 2-3$\sigma$ level~\cite{ELunghi}.  Given our success with $B_K$, we eventually aim to pursue $K\to\pi\pi$ decays with the mixed-action approach.

%%%%%%%%%%%%%%%

\end{document}